\documentclass[sigconf,xcolor,dvipsnames,usenames,screen]{acmart}

\usepackage[htt]{hyphenat}
\usepackage{cleveref}
\usepackage{enumitem}
\usepackage{algorithm}
\usepackage{algorithmicx}
\usepackage[noend]{algpseudocode}
\usepackage{tablefootnote}

\usepackage{xpatch}
\xpatchcmd\algorithmic{\leftmargin\labelwidth}{\leftmargin0.3\labelsep}{}{}
\algrenewcommand\alglinenumber[1]{\footnotesize #1}

\algrenewcommand\algorithmicprocedure{\textbf{function}}

\algblock{ParFor}{EndParFor}
\algnewcommand\algorithmicparfor{\textbf{parfor}}
\algnewcommand\algorithmicpardo{\textbf{}}
\algnewcommand\algorithmicendparfor{}
\algrenewtext{ParFor}[1]{\algorithmicparfor\ #1\ \algorithmicpardo}
\algrenewtext{EndParFor}{\algorithmicendparfor}
\algtext*{EndFor}
\algtext*{EndIf}
\algrenewcommand\algorithmicindent{0.8em}
\algrenewcommand\algorithmiccomment[1]{\hfill \textcolor{gray}{// #1}}

\usepackage{listings}
\usepackage{subfigure}

\DeclareFixedFont{\ttb}{T1}{txtt}{bx}{n}{12} 
\DeclareFixedFont{\ttm}{T1}{txtt}{m}{n}{12}  

\usepackage{color}
\definecolor{deepblue}{rgb}{0,0,0.5}
\definecolor{deepred}{rgb}{0.6,0,0}
\definecolor{deepgreen}{rgb}{0,0.5,0}

\usepackage{listings}

\newcommand\pythonstyle{\lstset{
language=Python,
basicstyle=\scriptsize\ttfamily,
morekeywords={self},              
keywordstyle=\scriptsize\ttfamily\color{deepblue},
emph={MyClass,__init__},          
emphstyle=\scriptsize\ttfamily\color{deepred},    
commentstyle=\scriptsize\color{gray}\ttfamily,
stringstyle=\scriptsize\ttfamily\color{deepgreen},
frame=none,                         
showstringspaces=false
}}

\lstnewenvironment{python}[1][]
{
\pythonstyle
\lstset{#1}
}
{}

\newcommand\pythoninline[1]{{\pythonstyle\lstinline!#1!}}
\usepackage{multirow}
\usepackage{tikz}

\usepackage{booktabs, tabularx}
\usepackage[skip=1pt]{caption}

\newcommand{\header}[1]{\vspace{1mm}\noindent\textbf{#1}.}

\newcommand{\headerl}[1]{\vspace{1mm}\noindent\textit{#1}.}
\usepackage{amsmath}
\usepackage{bm}

\DeclareMathOperator*{\argmin}{arg\,min}

\AtBeginDocument{%
  \providecommand\BibTeX{{%
    \normalfont B\kern-0.5em{\scshape i\kern-0.25em b}\kern-0.8em\TeX}}}

\setcopyright{none}
\acmYear{2025}
\acmDOI{}
\acmISBN{}
\acmConference[FAccTRec@RecSys'25]{FAccTRec@RecSys'25}{September 22--26, 2025}{Prague, Czech Republic}

\begin{document}

\title{Towards a Real-World Aligned Benchmark for~Unlearning~in~Recommender~Systems}

\author{%
Pierre Lubitzsch$^\dag$ $\,\,\,$ Olga Ovcharenko$^\dag$  $\,\,\,$ Hao Chen$^\dag$ $\,\,\,$ Maarten de Rijke$^\ddag$ $\,\,\,$  Sebastian Schelter$^\dag$}
\affiliation{%
  \institution{$^\dag$BIFOLD \& TU Berlin $\,\,\,$ $^\ddag$University of Amsterdam}
  \country{[lubitzsch,ovcharenko,hao.chen,schelter]@tu-berlin.de $\quad$ m.derijke@uva.nl}
}

\def\authors{Pierre Lubitzsch, Olga Ovcharenko, Hao Chen, Maarten de Rijke, and Sebastian Schelter}

\renewcommand{\shortauthors}{Lubitzsch et al.}
\begin{abstract}
Modern recommender systems heavily leverage user interaction data to deliver personalized experiences. However, relying on personal data presents challenges in adhering to privacy regulations, such as the GDPR's ``right to be forgotten''. Machine unlearning (MU) aims to address these challenges by enabling the efficient removal of specific training data from models post-training, without compromising model utility or leaving residual information. However, current benchmarks for unlearning in recommender systems---most notably CURE4Rec---fail to reflect real-world operational demands. They focus narrowly on collaborative filtering, overlook tasks like session-based and next-basket recommendation, simulate unrealistically large unlearning requests, and ignore critical efficiency constraints.
In this paper, we propose a set of design desiderata and research questions to guide the development of a more realistic benchmark for unlearning in recommender systems, with the goal of gathering feedback from the research community. Our benchmark proposal spans multiple recommendation tasks, includes domain-specific unlearning scenarios, and several unlearning algorithms---including ones adapted from a recent NeurIPS unlearning competition. Furthermore, we argue for an unlearning setup that reflects the sequential, time-sensitive nature of real-world deletion requests.
We also present a preliminary experiment in a next-basket recommendation setting based on our proposed desiderata and find that unlearning also works for sequential recommendation models, exposed to many small unlearning requests. In this case, we observe that a modification of a custom-designed unlearning algorithm for recommender systems outperforms general unlearning algorithms significantly, and that unlearning can be executed with a latency of only several seconds.
\end{abstract}




\maketitle

\section{Introduction}

Modern recommender systems rely heavily on user interaction data, such as clicks, ratings, and purchases to build personalized models.

\header{The right-to-be-forgotten and machine unlearning} This personalization enhances the user experience, but at the same time makes it challenging to adhere to privacy rights such as the ``right-to-be-forgotten'' from the GDPR~\cite{gdpr_article17} regulation in Europe and similar regulations adopted outside of Europe~\cite{ccpa_faq, DigiChina_AI_Rec}. These regulations mandate that responsible, ethical, and legally compliant AI systems give their users the right to request the timely deletion of their personal data from databases and models trained on them~\cite{stoyanovich2022responsible}. Being able to efficiently remove data from models is also important for security-related scenarios, e.g., to quickly react to spam interactions~\cite{amazonschoice} or unintended data leakage~\cite{carlini2019secret}.

On a technical level, the outlined challenges give rise to ``machine unlearning''~(MU), the problem of efficiently removing the influence of selected training data points from a machine learning model post training. After unlearning the data points there should be no information about these points in the model, it should ideally behave as if they had never been in the training set. Efficient machine unlearning is an active area of research~\cite{cao_2022, ginart2019makingaiforgetyou, izzo2021approximatedatadeletionmachine, schelter2021hedgecut, wang2022efficientlymaintainingbasketrecommendations, wu2020deltagradrapidretrainingmachine, Wu_2020}. The evaluation of unlearning algorithms is especially challenging since multiple aspects, such as the model utility after unlearning, the amount of information about the unlearned data still contained in the model (unlearning completeness), and the unlearning efficiency have to be considered simultaneously.

\header{Shortcomings of existing benchmarks for unlearning in recommender systems} We argue in \Cref{sec:design-shortcomings} that current unlearning benchmarks like CURE4Rec~\cite{chen2024cure4rec} do not sufficiently address the real-world demands of unlearning in recommender systems. They narrowly focus on collaborative filtering (CF)~\cite{bpr, ibcf, lightgcn, simrec, dccf}, overlooking other vital tasks, such as session-based recommendation (SBR)~\cite{hidasi2016sessionbasedrecommendationsrecurrentneural, sknn, gru4rec, srgnn, sasrec, narm} and next-basket recommendation (NBR)~\cite{rendle2010nbr, tifuknn, upcf, clea, dnntsp, sets2sets} that are common in e-commerce and frequently require unlearning capabilities. Moreover, these benchmarks simulate unlearning by deleting large chunks of data—typically 2–5\% of all interactions—whereas actual systems face continuous, small-scale deletion requests, such as when individual users withdraw consent for personalization. The choice of data to unlearn is also misrepresented: rather than being random, real scenarios often evolve around specific types of content, like interactions with sensitive items~\cite{schelter2023forget} or interactions from spammers~\cite{amazonschoice}. Operational efficiency is another overlooked factor; in practice, unlearning must be fast, inexpensive, and ideally executable directly on deployed models~\cite{schelter2021hedgecut}. The unlearning procedures in CURE4Rec are only an order of magnitude faster than retraining, which undermines real-time responsiveness. Finally, promising general-purpose unlearning algorithms, such as those highlighted in the NeurIPS 2023 competition~\cite{triantafillou2024makingprogressunlearningfindings}, remain underexplored for recommendation-specific applications.

\header{Towards a real-world aligned benchmark for unlearning in recommender systems} The goal of this paper is to propose the desiderata for a real-world aligned benchmark for unlearning in recommender systems, and to gather initial feedback.
In \Cref{sec:design-burst}, we formulate four extended research questions to address the shortcomings of existing benchmarks, and layout desiderata and design decisions for a new benchmark. In particular, we propose to cover NBR and SBR in addition to CF. For each recommendation task, we need different models, unlearning settings, and datasets to get insights about the unlearning performance for different data distributions. We consider various approximate unlearning algorithms for neural networks, one designed for recommendation models~\cite{li2023selectivecollaborativeinfluencefunction}, two adapted from the NeurIPS unlearning competition~\cite{triantafillou2024makingprogressunlearningfindings}, and three designed for graph neural networks (GNNs)~\cite{Wu_2023, kun_wu_2023, dong_2024} that are common architectures for recommender systems.
As recommendation models get fully retrained in the range between once per week (Amazon Prime~\cite{aws2025personalize}) and once per month (Tencent~\cite{tencentrecommendation2025}), it is unrealistic to unlearn a large chunk of data collected over a longer period of time. In contrast to existing benchmarks, we propose to issue a number of small unlearning requests in a sequential manner instead of unlearning a single forget set all at once, as unlearning requests are usually time-sensitive. We propose to evaluate the unlearned models based on retained model utility, unlearning completeness, and unlearning efficiency.

Moreover, we present a first preliminary experiment in~\Cref{sec:experiments}, designed according to our benchmark desiderata. Here, we evaluate the ability of three unlearning algorithms in a next-basket recommendation scenario, where we repeatedly unlearn the interactions with sensitive items (chosen in a domain-specific manner) from a neural model on an e-commerce dataset.

\header{Contributions} Our contributions are as follows:
\begin{itemize}[noitemsep, leftmargin=*]
  \item We discuss the shortcomings of existing benchmarks for unlearning in recommender systems~(\Cref{sec:design-shortcomings}).
  \item We propose a set of research questions and benchmark desiderata designed to address these shortcomings~(\Cref{sec:design-burst}).
  \item We present preliminary experimental results for an unlearning scenario in e-commerce, designed according to our benchmark desiderata. We find that we can successfully unlearn interactions with sensitive items for sequential recommendation models, exposed to many small unlearning requests. Furthermore, we observe that general unlearning algorithms are outperformed by a custom-designed unlearning algorithm for recommender systems with certain modifications, and that unlearning can be executed with a low latency of several seconds~(\Cref{sec:experiments}).
%
  \item We make the code for our experiments available at \url{https://github.com/pierre-lubitzsch/towards-unlearning-in-recsys}, based on the repository of \citeauthor{li2023basketrecommendationrealitycheck}~\cite{li2023basketrecommendationrealitycheck}. 
\end{itemize}

\section{Benchmark Proposal}
\label{sec:design}

We first elaborate on the shortcomings of existing unlearning benchmarks in \Cref{sec:design-shortcomings} and subsequently discuss how to address these shortcomings with a new benchmark in \Cref{sec:design-burst}.

\subsection{Shortcomings of Existing Benchmarks}
\label{sec:design-shortcomings}

We argue that current benchmarks for unlearning in recommender systems, such as CURE4Rec~\cite{chen2024cure4rec}, pose a valuable direction for academic research, but do not cover the characteristics and challenges of unlearning in real-world recommendation scenarios. We attribute this to the following reasons:


\headerl{Recommender systems are broader than collaborative filtering} In CURE4Rec~\cite{chen2024cure4rec} only collaborative filtering is considered, however the field of recommendation is much broader. Sequential recommendation tasks use temporal information to model evolving user preferences over time, with important applications in e-commerce and retail including SBR~\cite{hidasi2016sessionbasedrecommendationsrecurrentneural, sknn, gru4rec, srgnn, sasrec, narm} and NBR~\cite{rendle2010nbr, tifuknn, upcf, clea, dnntsp, sets2sets}. A comprehensive unlearning benchmark should also cover these tasks.

\begin{figure*}[t!]
\centering
\includegraphics[width=\textwidth]{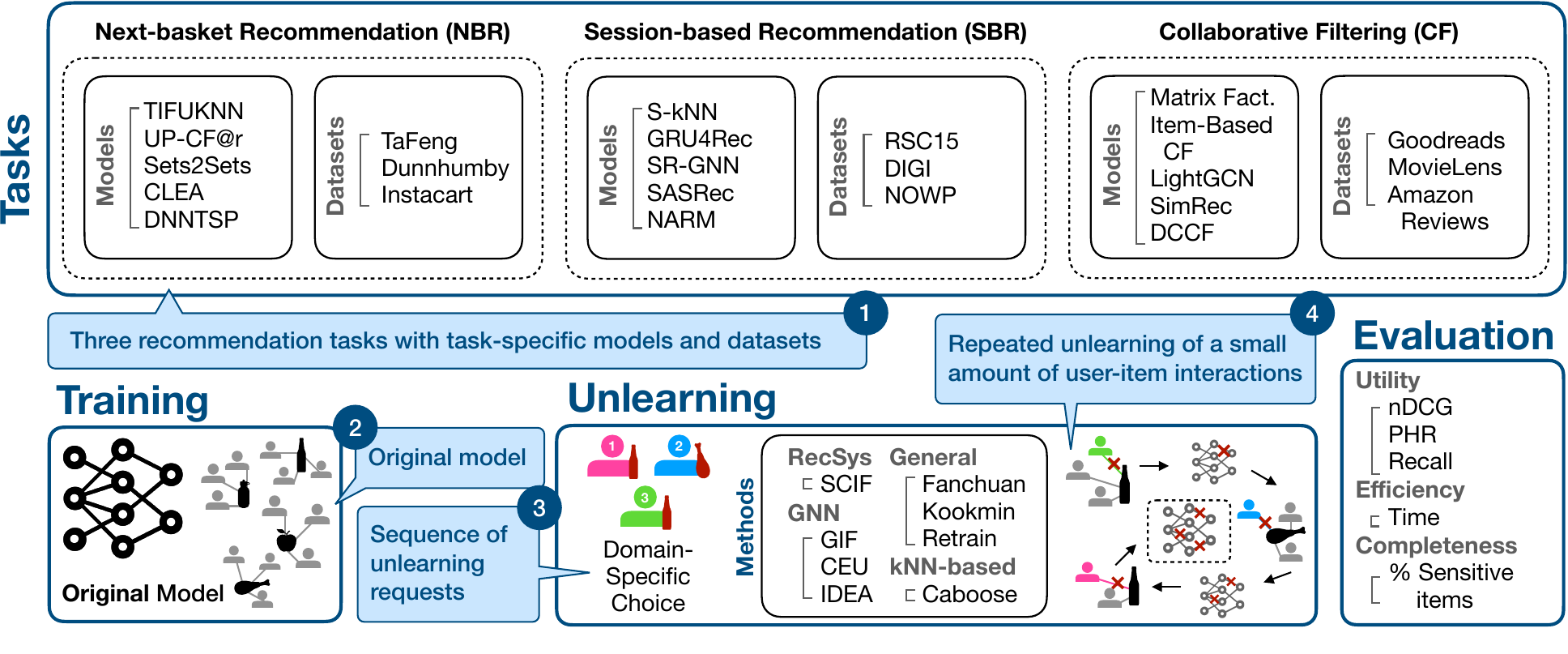}
\caption{Overview of the proposed benchmark.}
\label{fig:overview}
\end{figure*}

\headerl{Unlearning happens in small batches} In real-world applications, we expect many small unlearning requests to arrive over time, e.g., when users withdraw consent for personalization~\cite{schelter2021hedgecut} or when a security team asks to unlearn the interactions of certain spam users. It is important to handle many small repeated unlearning requests in a timely manner, ideally in a matter of seconds.
However, current benchmarks only evaluate a single, unrealistically large unlearning request. In the NeurIPS unlearning competition~\cite{triantafillou2024makingprogressunlearningfindings} and the CURE4Rec~\cite{chen2024cure4rec} benchmark, the data is unlearned at once, and the forget set size is 2\% and $\geq 5\%$ respectively. Big unlearning requests like this would never appear in industry because unlearning requests are incoming sequentially over time and for realistic user counts such percentages will not be reached before the model will be retrained either way. The European e-commerce company Bol, for example, has a recommender system for $>$13 million users~\cite{bol_about_2025} with a total of 2.3 billion total interactions~\cite{kersbergen2022serenade} as input. A forget set size of 5\% would correspond to unlearning the data of 650,000 users at once. It is hard to imagine why one would want to unlearn such a large number of users, and even if that were the case, the model would be retrained from scratch.

\headerl{Distribution of interactions to unlearn} In some domains, the choice of interactions to unlearn for a user follows a pattern. An example is the removal of interactions with sensitive items, for instance, alcoholic items for a user suffering from addiction~\cite{schelter2023forget,schelter2024snapcase}. Another area where the interactions follow a pattern is unlearning the interactions from malicious actors such as spammers, who try to manipulate the recommender system's behavior for a specific set of items~\cite{amazonschoice}. Current benchmarks do not evaluate domain-specific choices of the interactions to unlearn.

\headerl{Efficiency needs to be a first-class citizen} Unlearning needs to be executed fast, without utility loss. In CURE4Rec~\cite{chen2024cure4rec}, the runtime of unlearning is roughly one order of magnitude faster than retraining from scratch, which is not efficient enough for real-world use cases.

\headerl{Wide coverage of unlearning algorithms} The only approximate unlearning method in CURE4Rec is SCIF~\cite{li2023selectivecollaborativeinfluencefunction}, which conducts a single second-order parameter update. However, the 2023 NeurIPS unlearning competition~\cite{triantafillou2024makingprogressunlearningfindings} contains several high scoring first-order iterative approaches to unlearning, which also need to be evaluated for recommendation tasks.

\subsection{Research Questions and Benchmark~Desiderata}
\label{sec:design-burst}

 Motivated by the discussed shortcomings, we formulate the following research questions:
\begin{itemize}[noitemsep, leftmargin=*]
 \item {\em \textbf{RQ1} -- How well does unlearning perform for recommendation tasks beyond collaborative filtering?}
  \item {\em \textbf{RQ2} -- Do we need custom unlearning algorithms for recommender systems or are general algorithms for gradient-based unlearning enough?}
 \item {\em \textbf{RQ3} -- Can we successfully handle the repeated unlearning of small batches of interactions (chosen in a domain-specific manner)?}
 \item {\em \textbf{RQ4} -- Can we execute unlearning with sufficiently low latency to deploy it in real-world setups?}
\end{itemize}    

\noindent Based on these research questions, we discuss the desiderata and design of a real-world aligned benchmark for unlearning in recommender systems in this section. \Cref{fig:overview} visualizes the proposed benchmark.

\subsubsection{Choice of recommendation tasks and models}

To address {\em \textbf{RQ1}}, we propose to include a wide variety of recommendation tasks and models, in addition to classical collaborative filtering.

\header{Next-basket recommendation (NBR)} In this task, we are given a history of sets of item interactions of a user and aim to predict the set of items they will choose next. A common application of this is predicting the next shopping basket of a user in e-commerce or retail based on their past purchases. We plan to include two kNN-based models (\texttt{TIFUKNN}~\cite{tifuknn}, \texttt{UP-CF@r}~\cite{upcf}), the attention-based model~\texttt{Sets2Sets}~\cite{sets2sets}, \texttt{CLEA}~\cite{clea}, which applies constrastive learning and the GNN-based model \texttt{DNNTSP}~\cite{dnntsp}.

\header{Session-based recommendation (SBR)} This task evolves around predicting the next item that a user will interact with, based on a sequence of past interactions. Major applications are to predict the next product to show to a user in e-commerce or to predict the next song that a user will listen to on a music streaming platform. We plan to include the kNN-based model \texttt{S-kNN}~\cite{sknn}, two RNN-based models (\texttt{GRU4Rec}~\cite{gru4rec}, \texttt{NARM}~\cite{narm}), the attention-based model \texttt{SASRec}~\cite{sasrec} and the GNN-based model \texttt{SR-GNN}~\cite{srgnn}.

\header{Collaborative filtering (CF)} A classical task in recommender systems, already covered in CURE4Rec, where the goal is to predict the rating that a user would give to a particular unseen item, based on their past preferences. We include classical approaches such as \texttt{Matrix Factorization}~\cite{bpr} and item-based kNN~\cite{ibcf}, as well as three graph-based models (\texttt{LightGCN}~\cite{lightgcn}, \texttt{SimRec}~\cite{simrec}, \texttt{DCCF}~\cite{dccf}).

\subsubsection{Choice of datasets} 

We leverage a selection of widely used, publicly available datasets for our benchmark, as detailed in \Cref{tab:datasets}. The three NBR datasets are grocery shopping or retail related. Two of the three datasets for SBR are click datasets from e-commerce, while the third one is built from social media posts about songs listened to. For CF, we choose three common datasets containing ratings from users about items, extending the range of domains covered in CURE4Rec.

\begin{table*}[ht]
\centering
\setlength\tabcolsep{6.5pt}
\caption{Choice of datasets across various recommendation tasks for the proposed benchmark.}
\vspace{0.1cm}
\begin{tabular}{l l l r r r l}
\toprule
\textbf{Dataset} & \textbf{Domain} & \textbf{Task} & \textbf{\#Users/Sessions} & \textbf{\#Items} & \textbf{\#Interactions} & \textbf{Available at} \\
\midrule
TaFeng~\cite{tafeng} & Grocery shopping & NBR & 32,266 & 23,812 & 817,741 & \href{https://www.kaggle.com/datasets/chiranjivdas09/ta-feng-grocery-dataset}{Kaggle}\\
Dunnhumby~\cite{dunnhumby} & Retail shopping & NBR & 2,500 & 92,339 & 2,595,732 & \href{https://www.dunnhumby.com/source-files/}{dunnhumby.com}\\
Instacart~\cite{instacart} & Grocery shopping & NBR & 19,435 & 13,897 & 3,346,083 & \href{https://github.com/khanhnamle1994/instacart-orders/}{GitHub}\\
\midrule
RSC15~\cite{hidasi2016sessionbasedrecommendationsrecurrentneural} & E-commerce interactions & SBR & 9,249,729 & 52,739 & 33,003,944 & \href{https://www.kaggle.com/datasets/chadgostopp/recsys-challenge-2015/data}{Kaggle}\\
DIGI~\cite{digi} & E-commerce interactions & SBR& 310,325 & 156,813 & 1,235,381 & \href{https://competitions.codalab.org/competitions/11161}{Codalab}\\
NOWP~\cite{nowp} & Music on social media & SBR & 738,200 & 3,734,862 & 1,291,251,677 & \href{https://zenodo.org/records/2594483}{Zenodo}\\
\midrule
Goodreads~\cite{goodreads} & Book ratings & CF & 876,145 & 2,360,650 & 228,648,342 & \href{https://www.kaggle.com/datasets/pypiahmad/goodreads-book-reviews1}{Kaggle}\\
MovieLens~\cite{movielens20m} & Movie ratings & CF & 138,493 & 26,744 & 20,000,263 & \href{https://www.kaggle.com/datasets/grouplens/movielens-20m-dataset}{Kaggle}\\
Amazon Reviews~\cite{amazonreviews} & Product reviews & CF & 54,510,000 & 48,190,000 & 571,540,000 & \href{https://amazon-reviews-2023.github.io/}{GitHub}\\
\bottomrule
\end{tabular}
\label{tab:datasets}
\end{table*}

\subsubsection{Choice of unlearning algorithms}

To address {\em \textbf{RQ2}}, we propose to include recommendation-specific as well as general unlearning algorithms in the benchmark.
Unlearning algorithms not used during the preliminary experiments in this paper can be found in \Cref{app:unlearning-algorithms}.

\header{Custom unlearning algorithms for recommender systems}
We do not include general exact unlearning algorithms like RecEraser~\cite{chen2022recommendation} or UltraRE~\cite{li2023ultrare}, because their running time depends on the training set size, which is by design not efficient enough to handle a lot of queries with small forget sets. We focus on approximate unlearning algorithms with a time complexity linear in the forget set size. We propose to include the following algorithms:

\begin{itemize}[noitemsep,leftmargin=*]
    \item \texttt{SCIF}, a method to directly remove the influence of data in the forget set utilizing influence functions. The influence of the data in the forget set is calculated similarly to a Newton step, which additionally includes retain data to prevent catastrophic forgetting via 
$$
        \tilde{\theta}_{z \to \bar{z}} = \tilde{\theta} - H_{\tilde{\theta}}^{-1} \nabla_{\tilde{\theta}} (\frac{1}{2 + bs}(-\ell(z, \theta) + \ell(\bar{z}, \theta) + \sum_{i = 1}^{bs} \ell(z_i, \theta)))
$$
    where $z_i$ are retain samples, $z$ is an unlearning sample, $\bar{z}$ is the modified unlearning sample, $H$ is the Hessian, $bs$ is the number of retain samples used for the update, and $\tilde{\theta}$ is a subset of the parameters $\theta$ of the model~\cite{li2023selectivecollaborativeinfluencefunction}.
\end{itemize}

\header{General unlearning algorithms for gradient-based machine learning} We additionally propose to evaluate the following unlearning algorithms from the NeurIPS'23 unlearning challenge~\cite{triantafillou2024makingprogressunlearningfindings}, which have not specifically been designed for recommendation models. We select these particular approaches, as they scored the highest in the final ranking of the competition:

\begin{itemize}[noitemsep,leftmargin=*]
    \item \texttt{Fanchuan}: This method first operates on the forget set and learns to make the distribution for the output of the model on a unlearning sample uniform by minimizing the KL-Divergence between the output of the model on the unlearning samples and a uniform pseudo-label. In the next stage, the method varies between two modes for a certain number of epochs: in the first mode, we have a contrastive loss maximizing the distance between representations of samples in the forget set and samples in the retain set. In the other mode, we simply train the model on a subset of the retain data using the same procedure as training the model in a standard manner.
    \item \texttt{Kookmin}: This method takes the model and calculates the gradients of the loss w.r.t. the parameters of the model on the forget set and a subset of the retain set. This is done to identify parameters to reset, having a similar gradient in the forget and retain set. With this information, the method chooses the top $p \, |\theta|$ parameters $\hat{\theta}$ having the smallest absolute gradient difference for gradients from the forget and retain set: $\hat{\theta} = (\theta_{\bm{i}})_{\bm{i} \in \mathcal{J}}$ with $p \in (0, 1)$ being a hyperparameter.
    Here $\mathcal{J} = \argmin_{\mathcal{X} \subseteq \text{ind}(\theta), |\mathcal{X}| = \lfloor p |\theta| \rfloor } \sum_{\bm{i} \in \mathcal{X}} |(\nabla_{\theta} \ell(\theta, D_f))_{\bm{i}} - (\nabla_{\theta} \ell(\theta, B_r))_{\bm{i}}|$ is the index set choosing the subset of these parameters where $\text{ind}(t)$ references the index set of a tensor $t$ and $B_r$ is a batch of the retain set of size $O(|D_f|)$.
    These parameters will be re-initialized as if they were newly created, for example a parameter $w$ in a linear layer is re-initialized using uniform initialization~\cite{he2015delving} and other parameters $w$ are initialized by sampling from a centered normal distribution with low standard deviation: $w \sim \mathcal{N}(0, 0.02^2)$.
    After that, the model is trained on a subset of the retain set, where the parameters that were not re-initialized have a smaller learning rate by one order of magnitude.
\end{itemize}

\subsubsection{Evaluation metrics}

We discuss the dimensions and metrics to evaluate the performance of unlearning algorithms.

\header{Recommendation utility} We propose to leverage common ranking metrics such as Recall, Personalized hit ratio (PHR) and Normalized Discounted Cumulative Gain (nDCG).

\header{Unlearning completeness} Next to efficient unlearning and performance of the unlearned model, it is important to measure how much the model unlearned about the forget set. For exact unlearning methods, such evaluation is unnecessary because all information about the forget set is removed by definition. Measuring unlearning completeness for approximate methods is not trivial, because most approximate unlearning algorithms have no known theoretical guarantees (see the framework of $(\varepsilon, \delta)$-unlearning outlined in \Cref{sec:background}).
Without theoretical guarantees, there are several methods to measure unlearning empirically. Standard ones use the retrained model and calculate the distance in weight space or the distance in output distribution of the retrained and unlearned model.
Another approach to measuring unlearning completeness uses a hypothesis testing interpretation of unlearning with a fixed $\delta$ to get an estimated $\varepsilon$ for $(\varepsilon, \delta)$-unlearning \cite{triantafillou2024makingprogressunlearningfindings}.

\headerl{Sensitive item unlearning} A task specific evaluation method is sensitive item unlearning~\cite{schelter2023forget}. In this setting we want to remove items from a sensitive category for specific users, for example meat for users changing to a vegetarian diet, or alcohol for users suffering from addiction. For the evaluation, we count how many users still get sensitive items recommended after their unlearning request, in comparison to the recommendations of a retrained model. To evaluate recommendation utility we remove sensitive items from the test data of affected users.

\headerl{Removal of poisonous data} Another evaluation method involves removing poisonous data from malicious actors in session-based recommendation systems. These actors inject fake clicks on random items and items they want to promote, corrupting the training data and causing models to recommend irrelevant items. Unlearning success can be measured using standard performance metrics, as the goal is eliminating the negative impact of these fake interactions.

We propose to use both empirical setups to measure unlearning completeness and to construct corresponding experiments, which repeatedly unlearn small portions of the data, to account for {\em \textbf{RQ3}}.

\header{Unlearning efficiency} To account for {\em \textbf{RQ4}}, we propose to measure the time it takes to execute unlearning requests, as well as the required memory and hardware (e.g., if accelerator hardware is needed). Another important aspect is whether offline precomputation (independent of the choice of the forget set) is possible. In that case, we propose to report the precomputation time and the execution time of the unlearning algorithm separately. 


\section{Preliminary Results}
\label{sec:experiments}

In our preliminary experiment, we focus on the task of NBR and construct an experiment according to our desiderata from \Cref{sec:design-burst}.

\begin{table*}[t]
\centering
\setlength\tabcolsep{3.5pt}
\caption{Results for sensitive item unlearning on Sets2sets (Instacart dataset, averaged over five seeds). \texttt{Kookmin} and \texttt{Fanchuan} effectively remove sensitive items, with \texttt{Fanchuan} being fastest. \texttt{SCIF} performs badly in utility and completeness. However, when including norm clipping of the update step for \texttt{SCIF}, its performance drastically improves (\texttt{SCIF clip}). Bold/underlined entries show best/second-best performance across approximate algorithms for \% of requests.}
\vspace{0.1cm}
\setlength\extrarowheight{-1pt}
\begin{tabular}{ccc cc cc cc>{\raggedleft\arraybackslash}p{1.25cm}
                >{\raggedleft\arraybackslash}p{1.15cm}>{\raggedleft\arraybackslash}p{1.15cm}
                >{\raggedleft\arraybackslash}p{1.25cm}}
\toprule
\multirow{2}{*}{\textbf{Category}} &
\multirow{2}{*}{\textbf{Requests}} &
\multirow{2}{*}{\textbf{Algorithm}} &
\multicolumn{2}{c}{\textbf{Recall}} &
\multicolumn{2}{c}{\textbf{nDCG}} &
\multicolumn{2}{c}{\textbf{PHR}} &
\multicolumn{1}{c}{\textbf{Time per}} &
\multicolumn{2}{c}{\textbf{Sensitive items}} &
\multirow{2}{*}{\textbf{KL(R||U)}}\\
 &  &  &
 \textbf{@10} & \textbf{@20} &
 \textbf{@10} & \textbf{@20} &
 \textbf{@10} & \textbf{@20} &
 \multicolumn{1}{c}{\textbf{request (s)}} &
 \textbf{@10} & \textbf{@20}\\
\midrule
alcohol & 25\% & \texttt{Fanchuan} & {0.2071} & \underline{0.3026} & {0.2088} & {0.2379} & {0.6596} & {0.7725} & \textbf{6.31} & \textbf{0.00\%} & \textbf{0.00\%} & {3.7439} \\
alcohol & 50\% & \texttt{Fanchuan} & {0.2008} & {0.2909} & {0.2025} & {0.2294} & {0.6474} & {0.7632} & \textbf{6.31} & \textbf{0.00\%} & \textbf{0.00\%} & {4.2762} \\
alcohol & 75\% & \texttt{Fanchuan} & {0.2010} & {0.2891} & {0.2040} & {0.2296} & {0.6467} & {0.7615} & \textbf{6.31} & \textbf{0.00\%} & \underline{0.07\%} & {4.5580} \\
alcohol & 100\% & \texttt{Fanchuan} & {0.2022} & {0.2878} & {0.2061} & {0.2302} & {0.6493} & {0.7601} & \textbf{6.31} & \underline{0.04\%} & \underline{0.39\%} & {4.6788} \\
\midrule
alcohol & 25\% & \texttt{Kookmin} & \underline{0.2091} & {0.3005} & \textbf{0.2254} & \underline{0.2511} & \underline{0.6836} & \underline{0.7811} & \underline{46.35} & \textbf{0.00\%} & \textbf{0.00\%} & \underline{0.9800} \\
alcohol & 50\% & \texttt{Kookmin} & \underline{0.2078} & \underline{0.2984} & \textbf{0.2233} & \underline{0.2488} & \textbf{0.6816} & \underline{0.7792} & \underline{46.35} & \textbf{0.00\%} & \textbf{0.00\%} & \underline{1.0526} \\
alcohol & 75\% & \texttt{Kookmin} & \underline{0.2060} & \underline{0.2961} & \textbf{0.2213} & \underline{0.2468} & \textbf{0.6787} & \underline{0.7769} & \underline{46.35} & \textbf{0.00\%} & \underline{0.05\%} & \underline{1.1253} \\
alcohol & 100\% & \texttt{Kookmin} & \underline{0.2072} & \underline{0.2973} & \underline{0.2212} & \underline{0.2468} & \textbf{0.6807} & \underline{0.7776} & \underline{46.35} & \textbf{0.00\%} & \underline{0.06\%} & \underline{1.1269} \\
\midrule
alcohol & 25\% & \texttt{SCIF} & 0.1097 & 0.2080 & 0.0852 & 0.1244 & 0.4502 & 0.6626 & 55.40 & 25.09\% & 30.50\% & 17.3190 \\
alcohol & 50\% & \texttt{SCIF} & 0.1111 & 0.2092 & 0.0837 & 0.1229 & 0.4555 & 0.6646 & 55.40 & 17.86\% & 28.44\% & 19.8368 \\
alcohol & 75\% & \texttt{SCIF} & 0.1161 & 0.2106 & 0.0921 & 0.1289 & 0.4654 & 0.6631 & 55.40 & 17.66\% & 27.47\% & 33.4127 \\
alcohol & 100\% & \texttt{SCIF} & 0.1184 & 0.2131 & 0.0944 & 0.1313 & 0.4736 & 0.6686 & 55.40 & 17.75\% & 27.90\% & 38.3761 \\
\midrule
alcohol & 25\% & \texttt{SCIF clip} & \textbf{0.2175} & \textbf{0.3173} & \underline{0.2225} & \textbf{0.2528} & \textbf{0.6879} & \textbf{0.7888} & 95.67 & \textbf{0.00\%} & 0.37\% & \textbf{0.6082} \\
alcohol & 50\% & \texttt{SCIF clip} & \textbf{0.2145} & \textbf{0.3152} & \underline{0.2208} & \textbf{0.2518} & \underline{0.6767} & \textbf{0.7851} & 95.67 & \textbf{0.00\%} & 0.81\% & \textbf{0.7289} \\
alcohol & 75\% & \texttt{SCIF clip} & \textbf{0.2142} & \textbf{0.3145} & \underline{0.2222} & \textbf{0.2528} & \underline{0.6751} & \textbf{0.7840} & 95.67 & 0.05\% & 0.71\% & \textbf{0.8347} \\
alcohol & 100\% & \texttt{SCIF clip} & \textbf{0.2148} & \textbf{0.3146} & \textbf{0.2244} & \textbf{0.2546} & \underline{0.6754} & \textbf{0.7840} & 95.67 & \textbf{0.00\%} & 0.44\% & \textbf{0.8922} \\
\midrule
alcohol & 25\% & \texttt{Retrain} & 0.2220 & 0.3172 & 0.2295 & 0.2567 & 0.6968 & 0.7922 & 2977.40 & {0.00\%} & {0.00\%} & 0.0000 \\
alcohol & 50\% & \texttt{Retrain} & 0.2219 & 0.3172 & 0.2297 & 0.2570 & 0.6963 & 0.7925 & 3059.40 & {0.00\%} & {0.00\%} & 0.0000 \\
alcohol & 75\% & \texttt{Retrain} & 0.2216 & 0.3173 & 0.2299 & 0.2572 & 0.6971 & 0.7937 & 2939.20 & {0.00\%} & {0.00\%} & 0.0000 \\
alcohol & 100\% & \texttt{Retrain} & 0.2225 & 0.3178 & 0.2306 & 0.2577 & 0.6982 & 0.7937 & 3025.60 & {0.00\%} & {0.00\%} & 0.0000 \\
\midrule
meat & 25\% & \texttt{Fanchuan} & {0.2061} & {0.3006} & {0.2091} & {0.2374} & {0.6598} & {0.7716} & \textbf{6.00} & {0.36\%} & \underline{2.88\%} & {3.2123} \\
meat & 50\% & \texttt{Fanchuan} & {0.2045} & \underline{0.2984} & {0.2056} & {0.2340} & {0.6552} & {0.7695} & \textbf{6.00} & {0.31\%} & {4.80\%} & {3.4694} \\
meat & 75\% & \texttt{Fanchuan} & {0.2012} & {0.2923} & {0.2029} & {0.2302} & {0.6489} & {0.7644} & \textbf{6.00} & \textbf{0.00\%} & \textbf{0.52\%} & {3.8409} \\
meat & 100\% & \texttt{Fanchuan} & {0.2009} & {0.2886} & {0.2038} & {0.2293} & {0.6474} & {0.7613} & \textbf{6.00} & \underline{0.09\%} & \textbf{1.49\%} & {3.8590} \\
\midrule
meat & 25\% & \texttt{Kookmin} & \underline{0.2111} & \underline{0.3018} & \textbf{0.2276} & \underline{0.2528} & \underline{0.6877} & \underline{0.7825} & {46.15} & \textbf{0.12\%} & \textbf{1.31\%} & \underline{0.8741} \\
meat & 50\% & \texttt{Kookmin} & \underline{0.2078} & {0.2980} & \textbf{0.2239} & \underline{0.2491} & \textbf{0.6829} & \underline{0.7793} & {46.15} & \textbf{0.06\%} & \textbf{1.33\%} & \underline{0.9681} \\
meat & 75\% & \texttt{Kookmin} & \underline{0.2075} & \underline{0.2977} & \textbf{0.2229} & \underline{0.2484} & \textbf{0.6818} & \underline{0.7790} & {46.15} & {0.24\%} & {2.27\%} & \underline{1.0024} \\
meat & 100\% & \texttt{Kookmin} & \underline{0.2067} & \underline{0.2965} & \textbf{0.2223} & \underline{0.2476} & \textbf{0.6809} & \underline{0.7777} & {46.15} & {0.33\%} & {3.16\%} & \underline{1.0248} \\
\midrule
meat & 25\% & \texttt{SCIF} & 0.0965 & 0.1641 & 0.0819 & 0.1072 & 0.3680 & 0.5051 & \underline{38.50} & 49.96\% & 65.80\% & 25.7528 \\
meat & 50\% & \texttt{SCIF} & 0.0956 & 0.1634 & 0.0772 & 0.1030 & 0.3669 & 0.5040 & \underline{38.50} & 46.92\% & 61.09\% & 31.9693 \\
meat & 75\% & \texttt{SCIF} & 0.0958 & 0.1635 & 0.0772 & 0.1030 & 0.3677 & 0.5039 & \underline{38.50} & 46.74\% & 61.28\% & 34.0601 \\
meat & 100\% & \texttt{SCIF} & 0.0599 & 0.1087 & 0.0475 & 0.0665 & 0.2422 & 0.3561 & \underline{38.50} & 47.10\% & 72.02\% & 35.4680 \\
\midrule
meat & 25\% & \texttt{SCIF clip} & \textbf{0.2189} & \textbf{0.3182} & \underline{0.2240} & \textbf{0.2539} & \textbf{0.6879} & \textbf{0.7902} & 96.36 & \textbf{0.12\%} & 2.91\% & \textbf{0.5942} \\
meat & 50\% & \texttt{SCIF clip} & \textbf{0.2163} & \textbf{0.3169} & \underline{0.2213} & \textbf{0.2521} & \underline{0.6813} & \textbf{0.7868} & 96.36 & \underline{0.12\%} & \underline{1.95\%} & \textbf{0.6594} \\
meat & 75\% & \texttt{SCIF clip} & \textbf{0.2151} & \textbf{0.3157} & \underline{0.2210} & \textbf{0.2520} & \underline{0.6780} & \textbf{0.7854} & 96.36 & \underline{0.08\%} & \underline{1.54\%} & \textbf{0.7054} \\
meat & 100\% & \texttt{SCIF clip} & \textbf{0.2145} & \textbf{0.3151} & \underline{0.2213} & \textbf{0.2521} & \underline{0.6760} & \textbf{0.7841} & 96.36 & \textbf{0.03\%} & \underline{1.30\%} & \textbf{0.7466} \\
\midrule
meat & 25\% & \texttt{Retrain} & 0.2221 & 0.3183 & 0.2296 & 0.2573 & 0.6976 & 0.7935 & 3076.00 & {0.00\%} & 0.60\% & 0.0000 \\
meat & 50\% & \texttt{Retrain} & 0.2227 & 0.3195 & 0.2303 & 0.2582 & 0.6980 & 0.7948 & 3051.80 & {0.00\%} & 0.48\% & 0.0000 \\
meat & 75\% & \texttt{Retrain} & 0.2222 & 0.3185 & 0.2301 & 0.2578 & 0.6973 & 0.7942 & 3067.20 & 0.28\% & 3.18\% & 0.0000 \\
meat & 100\% & \texttt{Retrain} & 0.2222 & 0.3182 & 0.2301 & 0.2576 & 0.6975 & 0.7939 & 2601.80 & {0.00\%} & 0.82\% & 0.0000 \\
\bottomrule
\end{tabular}
\label{tab:sets2sets}
\end{table*}

\header{Experimental setup} We choose Sets2sets~\cite{sets2sets} as a model, which was among the best performing models in \citeauthor{li2023basketrecommendationrealitycheck}~\cite{li2023basketrecommendationrealitycheck}, and leverage the Instacart dataset, which contains transactions from grocery shopping. For unlearning, we evaluate the approximate neural unlearning techniques \texttt{SCIF}~\cite{li2023selectivecollaborativeinfluencefunction} and the two top approaches from the 2023 NeurIPS unlearning competition~\cite{triantafillou2024makingprogressunlearningfindings}. 

We evaluate the performance of the algorithms for sensitive item unlearning, with alcohol and meat as sensitive item categories. We train the models on our dataset with five different seeds. To construct a forget set for a sensitive category $c$, we randomly choose $0.1\%$ of the user-item interactions from items in baskets and sample only items from category $c$. 
This results in $\sim1000$ users per forget set for alcohol and $\sim650$ users per forget set for meat. For each seed we sample a different forget set. Let $D_f^{c,s}$ contain all forget requests sampled for category $c$ with seed $s$. We omit indices $c, s$ in further notation for readability. To unlearn $D_f$, we first fix an order of the users submitting a forget request: $(u_1, \dots, u_\ell)$.
Let $D$ be the initial training set, $D_f^i$ be unlearning request $i$ from $u_i$, $D_r^i = D \setminus \bigcup_{j \leq i} D_f^j$ be the retain set for request $i$, and $\theta_i$ be the weights of the model after unlearning $D_f^1, \dots, D_f^i$. We sequentially execute the unlearning algorithm on input $(\theta_{i - 1}, D_r^i, D_f^i)$ for $i \in  \{1, \dots, \ell\}$. We save $\theta_i$ at every quarter of the total forget set size $\ell$, so that we can evaluate the results after different amounts of unlearning requests.
For each (category, seed) combination, we unlearn one model per unlearning algorithm. We retrain models from scratch on $D_r^i$ for four checkpoints $i \in C := \{ \lceil \frac{\ell}{4} \rceil, \lceil \frac{2\ell}{4} \rceil, \lceil \frac{3\ell}{4} \rceil, \lceil \frac{4\ell}{4} \rceil \}$ and all (category, seed) combinations to compare the retrained models $\theta_i^r$ to the unlearned models $\theta_i$ for $i \in C$. Each user $u_j$ has one input sample $z$ consisting of their basket history. We use a temporal-order leave-one-out split to form a train/val/test split. When executing the unlearning algorithm on input $(\theta_{i - 1}, D_r^i, D_f^i)$, we unlearn all sensitive items from the baskets of $u_i$. 

Besides common utility metrics, we report the fraction of users that still receive recommendations for sensitive items after unlearning and the KL divergence between retrained and unlearned models for users who submitted unlearning requests, as a measure of unlearning completeness. To quantify efficiency, we report the average time to unlearn the interactions of a single user.  We conduct each experiment with a single NVIDIA A100 80GB PCIe GPU.

\headerl{Necessary adaptations for unlearning algorithms} The NBR setting requires us to adapt the unlearning algorithms. \texttt{Fanchuan} originally uses the whole retain set for the repair round. This would make the algorithm too inefficient in our case, and we limit the retain samples used per retain round to $10 \, |D_f^i|$ and only use 16 instead of 256 retain samples during the contrastive stage. \texttt{Kookmin} conducts a reset of 30\% of the model parameters in the setup for the NeurIPS competition; however, in our setting of sequential unlearning of small unlearning sets, it makes sense to lower this hyperparameter. We chose a value of $1\%$ of the parameters of all layers because 30\% made the model collapse due to shorter retain rounds, and $0.1\%$ rendered the updates negligible. We have 5 retain rounds and train on $32 \, |D_f^i|$ samples each round, instead of training 5 rounds on the full retain set. We adapt this to scale with the forget set for the sake of efficiency. Further studies on the best value of the reset percentage are needed. 
%
\texttt{SCIF} by default only updates the parameters for the user embeddings in CF models. However, in NBR, the information about a user is usually implicit within the basket history, and the chosen models do not contain user embeddings. Therefore, we make \texttt{SCIF} update the item embeddings, output linear layer, and a linear layer after attention for Sets2sets.

\header{Results and discussion} We list the results in \Cref{tab:sets2sets}. The performance of the adapted unlearning method \texttt{SCIF} is volatile. For some seeds the model performance collapsed, and for almost all users unlearning items were predicted. We attribute this observation to misguided update steps originating from an ill-conditioned Hessian approximation, which is confirmed by the large $\ell_2$-norm of the parameter update. Large update steps are rare but harm all subsequent updates due to the incremental nature of sequential unlearning.
With one seed, \texttt{SCIF} predicted 0\% items for the category meat, but the model utility was worse compared to other approaches. We measured the $\ell_2$-norm of the parameter update steps. In each training run, there were some update steps having norm $>100$ while the average norm over a run was close to $1$. A reason for that can be an ill-conditioned Hessian. Approaches to circumvent the big update steps are clipping update steps by norm or a bigger damping for the Hessian vector product calculation when update steps get too big. We decided to implement norm clipping to a maximum of $1$ for the update steps.
Let $\Delta \tilde{\theta} := - H_{\tilde{\theta}}^{-1} \nabla_{\tilde{\theta}} \left(\frac{1}{2 + bs}\left(-\ell(z, \theta) + \ell(\bar{z}, \theta) + \sum_{i = 1}^{bs} \ell(z_i, \theta)\right)\right)$.
We clip the update step norm to a maximum of $c \geq 0$:
$$
    \tilde{\theta} \leftarrow \tilde{\theta} - \min\left(1, \frac{c}{|| \Delta \tilde{\theta} ||_2}\right) \Delta \tilde{\theta},
$$
where $||\cdot||_2$ is the $\ell_2$-norm of the input tensor flattened into a 1-dimensional vector.\\
Preliminary results for this can be seen in rows with Algorithm \texttt{SCIF clip}. Due to time constraints $c = 1$ is chosen heuristically. Further experiments need to be done to find the best value of $c$ and to confirm the results for this seed. \texttt{SCIF clip} outperforms the other algorithms in model utility and unlearning completeness metrics. However, \texttt{SCIF clip} is an order of magnitude slower than \texttt{Fanchuan} and \texttt{Fanchuan} might perform better than \texttt{SCIF clip} given more time.\\
The general unlearning algorithms \texttt{Fanchuan} and \texttt{Kookmin} perform well. They achieve utility levels comparable to the retrained models, with a similarly effective reduction in sensitive item predictions. \texttt{Kookmin} performs slightly better in most cases. Furthermore, we observe that the average time to unlearn the interactions of a single user is in the range of six seconds for \texttt{Fanchuan}. From the KL divergence over the forget set, we can see that the predictions of \texttt{Kookmin} are always closer in distribution to the retrained model than the predictions from \texttt{Fanchuan}.

In summary, this experiment provides first evidence that our research questions from \Cref{sec:design-burst} can be positively answered with a new benchmark. The experiment indicates that unlearning also works for NBR models ({\em \textbf{RQ1}}) with many small unlearning requests ({\em \textbf{RQ3}}). We observe a case where a custom-designed unlearning algorithm for recommender systems outperforms general unlearning algorithms ({\em \textbf{RQ2}}). This only holds when applying update step norm clipping for the custom-designed unlearning algorithm. Without this modification, the performance drops significantly. Furthermore, unlearning algorithms can be executed with a low latency of several seconds only, which demonstrates the practical potential for deploying unlearning in real-world production systems~({\em \textbf{RQ4}}).
\section{Next Steps}
\label{sec:next}

Our plan for the future is to evolve our benchmark proposal (based on community feedback) into a comprehensive, fully specified, and openly available public benchmark, conduct all proposed experiments, and report on the results. We hope that our benchmark helps lay the foundation to deploy efficient unlearning for users in real-world systems.



\begin{acks}
    This research was (partially) supported by the Dutch Research Council (NWO), under project numbers 024.004.022, NWA.1389.20.\-183, and KICH3.LTP.20.006, and the European Union's Horizon Europe program under grant agreement No 101070212. All content represents the opinion of the authors, which is not necessarily shared or endorsed by their respective employers and/or sponsors.
\end{acks}

\balance

\bibliographystyle{ACM-Reference-Format}
\bibliography{rub}


\begin{thebibliography}{54}


\ifx \showCODEN    \undefined \def \showCODEN     #1{\unskip}     \fi
\ifx \showDOI      \undefined \def \showDOI       #1{#1}\fi
\ifx \showISBNx    \undefined \def \showISBNx     #1{\unskip}     \fi
\ifx \showISBNxiii \undefined \def \showISBNxiii  #1{\unskip}     \fi
\ifx \showISSN     \undefined \def \showISSN      #1{\unskip}     \fi
\ifx \showLCCN     \undefined \def \showLCCN      #1{\unskip}     \fi
\ifx \shownote     \undefined \def \shownote      #1{#1}          \fi
\ifx \showarticletitle \undefined \def \showarticletitle #1{#1}   \fi
\ifx \showURL      \undefined \def \showURL       {\relax}        \fi
\providecommand\bibfield[2]{#2}
\providecommand\bibinfo[2]{#2}
\providecommand\natexlab[1]{#1}
\providecommand\showeprint[2][]{arXiv:#2}

\bibitem[dig({[n.\,d.]})]%
        {digi}
 \bibinfo{year}{[n.\,d.]}\natexlab{}.
\newblock \bibinfo{title}{CIKM Cup 2016 Track 2: Personalized E-Commerce Search Challenge}.
\newblock \bibinfo{howpublished}{\url{https://competitions.codalab.org/competitions/11161\#learn_the_details-data2}}.
\newblock
\newblock
\shownote{Accessed: 2025-07-07}.


\bibitem[Dig(nd)]%
        {DigiChina_AI_Rec}
 \bibinfo{year}{[n.d.]}\natexlab{}.
\newblock \bibinfo{title}{Internet Information Service Algorithmic Recommendation Management Provisions}.
\newblock \bibinfo{howpublished}{\url{https://digichina.stanford.edu/work/translation-internet-information-servicealgorithmic-recommendation-management-provisions-opinon-seeking-draft/}}.
\newblock
\newblock
\shownote{Accessed: 2025-06-10}.


\bibitem[{Amazon Web Services}(2025)]%
        {aws2025personalize}
\bibfield{author}{\bibinfo{person}{{Amazon Web Services}}.} \bibinfo{year}{2025}\natexlab{}.
\newblock \bibinfo{booktitle}{\emph{Maintaining domain recommenders}}.
\newblock
\urldef\tempurl%
\url{https://docs.aws.amazon.com/personalize/latest/dg/maintaining-relevance.html#maintaining-domain-recommenders}
\showURL{%
\tempurl}
\newblock
\shownote{Accessed: 2025-06-30}.


\bibitem[{bol}(2025)]%
        {bol_about_2025}
\bibfield{author}{\bibinfo{person}{{bol}}.} \bibinfo{year}{2025}\natexlab{}.
\newblock \bibinfo{title}{About bol}.
\newblock
\newblock
\urldef\tempurl%
\url{https://over.bol.com/en/about-bol/}
\showURL{%
\tempurl}
\newblock
\shownote{Accessed: 2025-07-03}.


\bibitem[Buzzfeed(2019)]%
        {amazonschoice}
\bibfield{author}{\bibinfo{person}{Buzzfeed}.} \bibinfo{year}{2019}\natexlab{}.
\newblock \bibinfo{title}{“Amazon’s Choice” Does Not Necessarily Mean A Product Is Good}.
\newblock
\newblock
\urldef\tempurl%
\url{https://www.buzzfeednews.com/article/nicolenguyen/amazons-choice-bad-products}
\showURL{%
\tempurl}


\bibitem[{California Privacy Protection Agency}(nd)]%
        {ccpa_faq}
\bibfield{author}{\bibinfo{person}{{California Privacy Protection Agency}}.} \bibinfo{year}{[n.d.]}\natexlab{}.
\newblock \bibinfo{title}{California Consumer Privacy Act -- Frequently Asked Questions}.
\newblock \bibinfo{howpublished}{\url{https://cppa.ca.gov/faq.html\#faq_res_1}}.
\newblock
\newblock
\shownote{Accessed: 2025-06-10}.


\bibitem[Cao and Yang(2015)]%
        {cao_2022}
\bibfield{author}{\bibinfo{person}{Yinzhi Cao} {and} \bibinfo{person}{Junfeng Yang}.} \bibinfo{year}{2015}\natexlab{}.
\newblock \showarticletitle{Towards Making Systems Forget with Machine Unlearning}. In \bibinfo{booktitle}{\emph{2015 IEEE Symposium on Security and Privacy}}. \bibinfo{pages}{463--480}.
\newblock
\urldef\tempurl%
\url{https://doi.org/10.1109/SP.2015.35}
\showDOI{\tempurl}


\bibitem[Carlini et~al\mbox{.}(2019)]%
        {carlini2019secret}
\bibfield{author}{\bibinfo{person}{Nicholas Carlini}, \bibinfo{person}{Chang Liu}, \bibinfo{person}{{\'U}lfar Erlingsson}, \bibinfo{person}{Jernej Kos}, {and} \bibinfo{person}{Dawn Song}.} \bibinfo{year}{2019}\natexlab{}.
\newblock \showarticletitle{The secret sharer: Evaluating and testing unintended memorization in neural networks}. In \bibinfo{booktitle}{\emph{28th USENIX security symposium (USENIX security 19)}}. \bibinfo{pages}{267--284}.
\newblock


\bibitem[Chen et~al\mbox{.}(2022)]%
        {chen2022recommendation}
\bibfield{author}{\bibinfo{person}{Chong Chen}, \bibinfo{person}{Fei Sun}, \bibinfo{person}{Min Zhang}, {and} \bibinfo{person}{Bolin Ding}.} \bibinfo{year}{2022}\natexlab{}.
\newblock \showarticletitle{Recommendation unlearning}. In \bibinfo{booktitle}{\emph{Proceedings of the ACM web conference 2022}}. \bibinfo{pages}{2768--2777}.
\newblock


\bibitem[Chen et~al\mbox{.}(2024)]%
        {chen2024cure4rec}
\bibfield{author}{\bibinfo{person}{Chaochao Chen}, \bibinfo{person}{Jiaming Zhang}, \bibinfo{person}{Yizhao Zhang}, \bibinfo{person}{Li Zhang}, \bibinfo{person}{Lingjuan Lyu}, \bibinfo{person}{Yuyuan Li}, \bibinfo{person}{Biao Gong}, {and} \bibinfo{person}{Chenggang Yan}.} \bibinfo{year}{2024}\natexlab{}.
\newblock \showarticletitle{CURE4Rec: A benchmark for recommendation unlearning with deeper influence}.
\newblock \bibinfo{journal}{\emph{NeurIPS}} (\bibinfo{year}{2024}).
\newblock


\bibitem[Chen et~al\mbox{.}(2025)]%
        {tencentrecommendation2025}
\bibfield{author}{\bibinfo{person}{Zihao Chen}, \bibinfo{person}{Chenyang Zhang}, \bibinfo{person}{Chen Xu}, \bibinfo{person}{Zhao Zhang}, \bibinfo{person}{Jiaqiang Wang}, \bibinfo{person}{Weining Qian}, {and} \bibinfo{person}{Aoying Zhou}.} \bibinfo{year}{2025}\natexlab{}.
\newblock \showarticletitle{Scheduling Data Processing Pipelines for Incremental Training on MLP-based Recommendation Models}. In \bibinfo{booktitle}{\emph{Companion of the 2025 International Conference on Management of Data}} (Berlin, Germany) \emph{(\bibinfo{series}{SIGMOD/PODS '25})}. \bibinfo{publisher}{Association for Computing Machinery}, \bibinfo{address}{New York, NY, USA}, \bibinfo{pages}{350–363}.
\newblock
\showISBNx{9798400715648}
\urldef\tempurl%
\url{https://doi.org/10.1145/3722212.3724454}
\showDOI{\tempurl}


\bibitem[Das(2025)]%
        {tafeng}
\bibfield{author}{\bibinfo{person}{Chiranjiv Das}.} \bibinfo{year}{2025}\natexlab{}.
\newblock \bibinfo{title}{Ta-Feng Grocery Dataset}.
\newblock \bibinfo{howpublished}{\url{https://www.kaggle.com/datasets/chiranjivdas09/ta-feng-grocery-dataset}}.
\newblock
\newblock
\shownote{Accessed: 2025-07-07}.


\bibitem[Dong et~al\mbox{.}(2024)]%
        {dong_2024}
\bibfield{author}{\bibinfo{person}{Yushun Dong}, \bibinfo{person}{Binchi Zhang}, \bibinfo{person}{Zhenyu Lei}, \bibinfo{person}{Na Zou}, {and} \bibinfo{person}{Jundong Li}.} \bibinfo{year}{2024}\natexlab{}.
\newblock \showarticletitle{IDEA: A Flexible Framework of Certified Unlearning for Graph Neural Networks}. In \bibinfo{booktitle}{\emph{Proceedings of the 30th ACM SIGKDD Conference on Knowledge Discovery and Data Mining}} (Barcelona, Spain) \emph{(\bibinfo{series}{KDD '24})}. \bibinfo{publisher}{Association for Computing Machinery}, \bibinfo{address}{New York, NY, USA}, \bibinfo{pages}{621–630}.
\newblock
\showISBNx{9798400704901}
\urldef\tempurl%
\url{https://doi.org/10.1145/3637528.3671744}
\showDOI{\tempurl}


\bibitem[{dunnhumby}(2025)]%
        {dunnhumby}
\bibfield{author}{\bibinfo{person}{{dunnhumby}}.} \bibinfo{year}{2025}\natexlab{}.
\newblock \bibinfo{title}{Source Files Datasets}.
\newblock \bibinfo{howpublished}{\url{https://www.dunnhumby.com/source-files/}}.
\newblock
\newblock
\shownote{Accessed: 2025-07-07}.


\bibitem[Faggioli et~al\mbox{.}(2020)]%
        {upcf}
\bibfield{author}{\bibinfo{person}{Guglielmo Faggioli}, \bibinfo{person}{Mirko Polato}, {and} \bibinfo{person}{Fabio Aiolli}.} \bibinfo{year}{2020}\natexlab{}.
\newblock \showarticletitle{Recency Aware Collaborative Filtering for Next Basket Recommendation}. In \bibinfo{booktitle}{\emph{Proceedings of the 28th ACM Conference on User Modeling, Adaptation and Personalization}} (Genoa, Italy) \emph{(\bibinfo{series}{UMAP '20})}. \bibinfo{publisher}{Association for Computing Machinery}, \bibinfo{address}{New York, NY, USA}, \bibinfo{pages}{80–87}.
\newblock
\showISBNx{9781450368612}
\urldef\tempurl%
\url{https://doi.org/10.1145/3340631.3394850}
\showDOI{\tempurl}


\bibitem[{GDPR.eu}(nd)]%
        {gdpr_article17}
\bibfield{author}{\bibinfo{person}{{GDPR.eu}}.} \bibinfo{year}{[n.d.]}\natexlab{}.
\newblock \bibinfo{title}{Article 17: Right to be forgotten}.
\newblock \bibinfo{howpublished}{\url{https://gdpr.eu/article-17-rightto-be-forgotten}}.
\newblock
\newblock
\shownote{Accessed: 2025-06-10}.


\bibitem[Ginart et~al\mbox{.}(2019)]%
        {ginart2019makingaiforgetyou}
\bibfield{author}{\bibinfo{person}{Antonio Ginart}, \bibinfo{person}{Melody~Y. Guan}, \bibinfo{person}{Gregory Valiant}, {and} \bibinfo{person}{James Zou}.} \bibinfo{year}{2019}\natexlab{}.
\newblock \bibinfo{title}{Making AI Forget You: Data Deletion in Machine Learning}.
\newblock
\newblock
\showeprint[arxiv]{1907.05012}~[cs.LG]
\urldef\tempurl%
\url{https://arxiv.org/abs/1907.05012}
\showURL{%
\tempurl}


\bibitem[Harper and Konstan(2015)]%
        {movielens20m}
\bibfield{author}{\bibinfo{person}{F.~Maxwell Harper} {and} \bibinfo{person}{Joseph~A. Konstan}.} \bibinfo{year}{2015}\natexlab{}.
\newblock \showarticletitle{The MovieLens Datasets: History and Context}.
\newblock \bibinfo{journal}{\emph{ACM Transactions on Interactive Intelligent Systems (TiiS)}} \bibinfo{volume}{5}, \bibinfo{number}{4} (\bibinfo{year}{2015}), \bibinfo{pages}{19:1--19:19}.
\newblock
\urldef\tempurl%
\url{https://doi.org/10.1145/2827872}
\showDOI{\tempurl}


\bibitem[He et~al\mbox{.}(2015)]%
        {he2015delving}
\bibfield{author}{\bibinfo{person}{Kaiming He}, \bibinfo{person}{Xiangyu Zhang}, \bibinfo{person}{Shaoqing Ren}, {and} \bibinfo{person}{Jian Sun}.} \bibinfo{year}{2015}\natexlab{}.
\newblock \showarticletitle{Delving Deep into Rectifiers: Surpassing Human‐Level Performance on ImageNet Classification}. In \bibinfo{booktitle}{\emph{Proceedings of the IEEE International Conference on Computer Vision (ICCV)}}. \bibinfo{pages}{1026--1034}.
\newblock
\urldef\tempurl%
\url{https://arxiv.org/abs/1502.01852}
\showURL{%
\tempurl}


\bibitem[He et~al\mbox{.}(2020)]%
        {lightgcn}
\bibfield{author}{\bibinfo{person}{Xiangnan He}, \bibinfo{person}{Kuan Deng}, \bibinfo{person}{Xiang Wang}, \bibinfo{person}{Yan Li}, \bibinfo{person}{YongDong Zhang}, {and} \bibinfo{person}{Meng Wang}.} \bibinfo{year}{2020}\natexlab{}.
\newblock \showarticletitle{LightGCN: Simplifying and Powering Graph Convolution Network for Recommendation}. In \bibinfo{booktitle}{\emph{Proceedings of the 43rd International ACM SIGIR Conference on Research and Development in Information Retrieval}} (Virtual Event, China) \emph{(\bibinfo{series}{SIGIR '20})}. \bibinfo{publisher}{Association for Computing Machinery}, \bibinfo{address}{New York, NY, USA}, \bibinfo{pages}{639–648}.
\newblock
\showISBNx{9781450380164}
\urldef\tempurl%
\url{https://doi.org/10.1145/3397271.3401063}
\showDOI{\tempurl}


\bibitem[Hidasi et~al\mbox{.}(2016)]%
        {hidasi2016sessionbasedrecommendationsrecurrentneural}
\bibfield{author}{\bibinfo{person}{Balázs Hidasi}, \bibinfo{person}{Alexandros Karatzoglou}, \bibinfo{person}{Linas Baltrunas}, {and} \bibinfo{person}{Domonkos Tikk}.} \bibinfo{year}{2016}\natexlab{}.
\newblock \bibinfo{title}{Session-based Recommendations with Recurrent Neural Networks}.
\newblock
\newblock
\showeprint[arxiv]{1511.06939}~[cs.LG]
\urldef\tempurl%
\url{https://arxiv.org/abs/1511.06939}
\showURL{%
\tempurl}


\bibitem[Hou et~al\mbox{.}(2024)]%
        {amazonreviews}
\bibfield{author}{\bibinfo{person}{Yupeng Hou}, \bibinfo{person}{Jiacheng Li}, \bibinfo{person}{Zhankui He}, \bibinfo{person}{An Yan}, \bibinfo{person}{Xiusi Chen}, {and} \bibinfo{person}{Julian McAuley}.} \bibinfo{year}{2024}\natexlab{}.
\newblock \showarticletitle{Bridging Language and Items for Retrieval and Recommendation}.
\newblock \bibinfo{journal}{\emph{arXiv preprint arXiv:2403.03952}} (\bibinfo{year}{2024}).
\newblock


\bibitem[Hu and He(2019)]%
        {sets2sets}
\bibfield{author}{\bibinfo{person}{Haoji Hu} {and} \bibinfo{person}{Xiangnan He}.} \bibinfo{year}{2019}\natexlab{}.
\newblock \showarticletitle{Sets2Sets: Learning from Sequential Sets with Neural Networks}. In \bibinfo{booktitle}{\emph{Proceedings of the 25th ACM SIGKDD International Conference on Knowledge Discovery \& Data Mining}} (Anchorage, AK, USA) \emph{(\bibinfo{series}{KDD '19})}. \bibinfo{publisher}{Association for Computing Machinery}, \bibinfo{address}{New York, NY, USA}, \bibinfo{pages}{1491–1499}.
\newblock
\showISBNx{9781450362016}
\urldef\tempurl%
\url{https://doi.org/10.1145/3292500.3330979}
\showDOI{\tempurl}


\bibitem[Hu et~al\mbox{.}(2020)]%
        {tifuknn}
\bibfield{author}{\bibinfo{person}{Haoji Hu}, \bibinfo{person}{Xiangnan He}, \bibinfo{person}{Jinyang Gao}, {and} \bibinfo{person}{Zhi-Li Zhang}.} \bibinfo{year}{2020}\natexlab{}.
\newblock \showarticletitle{Modeling Personalized Item Frequency Information for Next-basket Recommendation}. In \bibinfo{booktitle}{\emph{Proceedings of the 43rd International ACM SIGIR Conference on Research and Development in Information Retrieval}} (Virtual Event, China) \emph{(\bibinfo{series}{SIGIR '20})}. \bibinfo{publisher}{Association for Computing Machinery}, \bibinfo{address}{New York, NY, USA}, \bibinfo{pages}{1071–1080}.
\newblock
\showISBNx{9781450380164}
\urldef\tempurl%
\url{https://doi.org/10.1145/3397271.3401066}
\showDOI{\tempurl}


\bibitem[Izzo et~al\mbox{.}(2021)]%
        {izzo2021approximatedatadeletionmachine}
\bibfield{author}{\bibinfo{person}{Zachary Izzo}, \bibinfo{person}{Mary~Anne Smart}, \bibinfo{person}{Kamalika Chaudhuri}, {and} \bibinfo{person}{James Zou}.} \bibinfo{year}{2021}\natexlab{}.
\newblock \bibinfo{title}{Approximate Data Deletion from Machine Learning Models}.
\newblock
\newblock
\showeprint[arxiv]{2002.10077}~[cs.LG]
\urldef\tempurl%
\url{https://arxiv.org/abs/2002.10077}
\showURL{%
\tempurl}


\bibitem[Jannach and Ludewig(2017)]%
        {sknn}
\bibfield{author}{\bibinfo{person}{Dietmar Jannach} {and} \bibinfo{person}{Malte Ludewig}.} \bibinfo{year}{2017}\natexlab{}.
\newblock \showarticletitle{When Recurrent Neural Networks meet the Neighborhood for Session-Based Recommendation}. In \bibinfo{booktitle}{\emph{Proceedings of the Eleventh ACM Conference on Recommender Systems}} (Como, Italy) \emph{(\bibinfo{series}{RecSys '17})}. \bibinfo{publisher}{Association for Computing Machinery}, \bibinfo{address}{New York, NY, USA}, \bibinfo{pages}{306–310}.
\newblock
\showISBNx{9781450346528}
\urldef\tempurl%
\url{https://doi.org/10.1145/3109859.3109872}
\showDOI{\tempurl}


\bibitem[Kang and McAuley(2018)]%
        {sasrec}
\bibfield{author}{\bibinfo{person}{Wang-Cheng Kang} {and} \bibinfo{person}{Julian McAuley}.} \bibinfo{year}{2018}\natexlab{}.
\newblock \showarticletitle{Self-Attentive Sequential Recommendation}. In \bibinfo{booktitle}{\emph{2018 IEEE International Conference on Data Mining (ICDM)}}. \bibinfo{pages}{197--206}.
\newblock
\urldef\tempurl%
\url{https://doi.org/10.1109/ICDM.2018.00035}
\showDOI{\tempurl}


\bibitem[Kersbergen et~al\mbox{.}(2022)]%
        {kersbergen2022serenade}
\bibfield{author}{\bibinfo{person}{Barrie Kersbergen}, \bibinfo{person}{Olivier Sprangers}, {and} \bibinfo{person}{Sebastian Schelter}.} \bibinfo{year}{2022}\natexlab{}.
\newblock \showarticletitle{Serenade-low-latency session-based recommendation in e-commerce at scale}. In \bibinfo{booktitle}{\emph{Proceedings of the 2022 International Conference on Management of Data}}. \bibinfo{pages}{150--159}.
\newblock


\bibitem[Le(2025)]%
        {instacart}
\bibfield{author}{\bibinfo{person}{Khanh~Nam Le}.} \bibinfo{year}{2025}\natexlab{}.
\newblock \bibinfo{title}{Instacart Market Basket Analysis Dataset}.
\newblock \bibinfo{howpublished}{\url{https://github.com/khanhnamle1994/instacart-orders/tree/master/data}}.
\newblock
\newblock
\shownote{Accessed: 2025-07-07}.


\bibitem[Li et~al\mbox{.}(2017)]%
        {narm}
\bibfield{author}{\bibinfo{person}{Jing Li}, \bibinfo{person}{Pengjie Ren}, \bibinfo{person}{Zhumin Chen}, \bibinfo{person}{Zhaochun Ren}, \bibinfo{person}{Tao Lian}, {and} \bibinfo{person}{Jun Ma}.} \bibinfo{year}{2017}\natexlab{}.
\newblock \showarticletitle{Neural Attentive Session-based Recommendation}. In \bibinfo{booktitle}{\emph{Proceedings of the 2017 ACM on Conference on Information and Knowledge Management}} (Singapore, Singapore) \emph{(\bibinfo{series}{CIKM '17})}. \bibinfo{publisher}{Association for Computing Machinery}, \bibinfo{address}{New York, NY, USA}, \bibinfo{pages}{1419–1428}.
\newblock
\showISBNx{9781450349185}
\urldef\tempurl%
\url{https://doi.org/10.1145/3132847.3132926}
\showDOI{\tempurl}


\bibitem[Li et~al\mbox{.}(2023c)]%
        {li2023basketrecommendationrealitycheck}
\bibfield{author}{\bibinfo{person}{Ming Li}, \bibinfo{person}{Sami Jullien}, \bibinfo{person}{Mozhdeh Ariannezhad}, {and} \bibinfo{person}{Maarten de Rijke}.} \bibinfo{year}{2023}\natexlab{c}.
\newblock \bibinfo{title}{A Next Basket Recommendation Reality Check}.
\newblock
\newblock
\showeprint[arxiv]{2109.14233}~[cs.IR]
\urldef\tempurl%
\url{https://arxiv.org/abs/2109.14233}
\showURL{%
\tempurl}


\bibitem[Li et~al\mbox{.}(2023a)]%
        {li2023ultrare}
\bibfield{author}{\bibinfo{person}{Yuyuan Li}, \bibinfo{person}{Chaochao Chen}, \bibinfo{person}{Yizhao Zhang}, \bibinfo{person}{Weiming Liu}, \bibinfo{person}{Lingjuan Lyu}, \bibinfo{person}{Xiaolin Zheng}, \bibinfo{person}{Dan Meng}, {and} \bibinfo{person}{Jun Wang}.} \bibinfo{year}{2023}\natexlab{a}.
\newblock \showarticletitle{Ultrare: Enhancing receraser for recommendation unlearning via error decomposition}.
\newblock \bibinfo{journal}{\emph{Advances in Neural Information Processing Systems}}  \bibinfo{volume}{36} (\bibinfo{year}{2023}), \bibinfo{pages}{12611--12625}.
\newblock


\bibitem[Li et~al\mbox{.}(2023b)]%
        {li2023selectivecollaborativeinfluencefunction}
\bibfield{author}{\bibinfo{person}{Yuyuan Li}, \bibinfo{person}{Chaochao Chen}, \bibinfo{person}{Xiaolin Zheng}, \bibinfo{person}{Yizhao Zhang}, \bibinfo{person}{Biao Gong}, {and} \bibinfo{person}{Jun Wang}.} \bibinfo{year}{2023}\natexlab{b}.
\newblock \bibinfo{title}{Selective and Collaborative Influence Function for Efficient Recommendation Unlearning}.
\newblock
\newblock
\showeprint[arxiv]{2304.10199}~[cs.IR]
\urldef\tempurl%
\url{https://arxiv.org/abs/2304.10199}
\showURL{%
\tempurl}


\bibitem[pypiahmad(2025)]%
        {goodreads}
\bibfield{author}{\bibinfo{person}{pypiahmad}.} \bibinfo{year}{2025}\natexlab{}.
\newblock \bibinfo{title}{Goodreads Book Reviews Dataset}.
\newblock \bibinfo{howpublished}{\url{https://www.kaggle.com/datasets/pypiahmad/goodreads-book-reviews1}}.
\newblock
\newblock
\shownote{Accessed: 2025-07-07}.


\bibitem[Qin et~al\mbox{.}(2021)]%
        {clea}
\bibfield{author}{\bibinfo{person}{Yuqi Qin}, \bibinfo{person}{Pengfei Wang}, {and} \bibinfo{person}{Chenliang Li}.} \bibinfo{year}{2021}\natexlab{}.
\newblock \showarticletitle{The World is Binary: Contrastive Learning for Denoising Next Basket Recommendation}. In \bibinfo{booktitle}{\emph{Proceedings of the 44th International ACM SIGIR Conference on Research and Development in Information Retrieval}} (Virtual Event, Canada) \emph{(\bibinfo{series}{SIGIR '21})}. \bibinfo{publisher}{Association for Computing Machinery}, \bibinfo{address}{New York, NY, USA}, \bibinfo{pages}{859–868}.
\newblock
\showISBNx{9781450380379}
\urldef\tempurl%
\url{https://doi.org/10.1145/3404835.3462836}
\showDOI{\tempurl}


\bibitem[Ren et~al\mbox{.}(2023)]%
        {dccf}
\bibfield{author}{\bibinfo{person}{Xubin Ren}, \bibinfo{person}{Lianghao Xia}, \bibinfo{person}{Jiashu Zhao}, \bibinfo{person}{Dawei Yin}, {and} \bibinfo{person}{Chao Huang}.} \bibinfo{year}{2023}\natexlab{}.
\newblock \showarticletitle{Disentangled Contrastive Collaborative Filtering}. In \bibinfo{booktitle}{\emph{Proceedings of the 46th International ACM SIGIR Conference on Research and Development in Information Retrieval}}. \bibinfo{publisher}{ACM}, \bibinfo{pages}{1137–1146}.
\newblock
\urldef\tempurl%
\url{https://doi.org/10.1145/3539618.3591665}
\showDOI{\tempurl}


\bibitem[Rendle et~al\mbox{.}(2009)]%
        {bpr}
\bibfield{author}{\bibinfo{person}{Steffen Rendle}, \bibinfo{person}{Christoph Freudenthaler}, \bibinfo{person}{Zeno Gantner}, {and} \bibinfo{person}{Lars Schmidt-Thieme}.} \bibinfo{year}{2009}\natexlab{}.
\newblock \showarticletitle{BPR: Bayesian personalized ranking from implicit feedback}. In \bibinfo{booktitle}{\emph{Proceedings of the Twenty-Fifth Conference on Uncertainty in Artificial Intelligence}} (Montreal, Quebec, Canada) \emph{(\bibinfo{series}{UAI '09})}. \bibinfo{publisher}{AUAI Press}, \bibinfo{address}{Arlington, Virginia, USA}, \bibinfo{pages}{452–461}.
\newblock
\showISBNx{9780974903958}


\bibitem[Rendle et~al\mbox{.}(2010)]%
        {rendle2010nbr}
\bibfield{author}{\bibinfo{person}{Steffen Rendle}, \bibinfo{person}{Christoph Freudenthaler}, {and} \bibinfo{person}{Lars Schmidt-Thieme}.} \bibinfo{year}{2010}\natexlab{}.
\newblock \showarticletitle{Factorizing personalized Markov chains for next-basket recommendation}. In \bibinfo{booktitle}{\emph{Proceedings of the 19th International Conference on World Wide Web}} (Raleigh, North Carolina, USA) \emph{(\bibinfo{series}{WWW '10})}. \bibinfo{publisher}{Association for Computing Machinery}, \bibinfo{address}{New York, NY, USA}, \bibinfo{pages}{811–820}.
\newblock
\showISBNx{9781605587998}
\urldef\tempurl%
\url{https://doi.org/10.1145/1772690.1772773}
\showDOI{\tempurl}


\bibitem[Sarwar et~al\mbox{.}(2001)]%
        {ibcf}
\bibfield{author}{\bibinfo{person}{Badrul Sarwar}, \bibinfo{person}{George Karypis}, \bibinfo{person}{Joseph Konstan}, {and} \bibinfo{person}{John Riedl}.} \bibinfo{year}{2001}\natexlab{}.
\newblock \showarticletitle{Item-based collaborative filtering recommendation algorithms}. In \bibinfo{booktitle}{\emph{Proceedings of the 10th International Conference on World Wide Web}} (Hong Kong, Hong Kong) \emph{(\bibinfo{series}{WWW '01})}. \bibinfo{publisher}{Association for Computing Machinery}, \bibinfo{address}{New York, NY, USA}, \bibinfo{pages}{285–295}.
\newblock
\showISBNx{1581133480}
\urldef\tempurl%
\url{https://doi.org/10.1145/371920.372071}
\showDOI{\tempurl}


\bibitem[Schelter et~al\mbox{.}(2023)]%
        {schelter2023forget}
\bibfield{author}{\bibinfo{person}{Sebastian Schelter}, \bibinfo{person}{Mozhdeh Ariannezhad}, {and} \bibinfo{person}{Maarten de Rijke}.} \bibinfo{year}{2023}\natexlab{}.
\newblock \showarticletitle{Forget me now: Fast and exact unlearning in neighborhood-based recommendation}. In \bibinfo{booktitle}{\emph{Proceedings of the 46th International ACM SIGIR Conference on Research and Development in Information Retrieval}}. \bibinfo{pages}{2011--2015}.
\newblock


\bibitem[Schelter et~al\mbox{.}(2024)]%
        {schelter2024snapcase}
\bibfield{author}{\bibinfo{person}{Sebastian Schelter}, \bibinfo{person}{Stefan Grafberger}, {and} \bibinfo{person}{Maarten de Rijke}.} \bibinfo{year}{2024}\natexlab{}.
\newblock \showarticletitle{Snapcase-Regain Control over Your Predictions with Low-Latency Machine Unlearning}.
\newblock \bibinfo{journal}{\emph{Proceedings of the VLDB Endowment}} \bibinfo{volume}{17}, \bibinfo{number}{12} (\bibinfo{year}{2024}), \bibinfo{pages}{4273--4276}.
\newblock


\bibitem[Schelter et~al\mbox{.}(2021)]%
        {schelter2021hedgecut}
\bibfield{author}{\bibinfo{person}{Sebastian Schelter}, \bibinfo{person}{Stefan Grafberger}, {and} \bibinfo{person}{Ted Dunning}.} \bibinfo{year}{2021}\natexlab{}.
\newblock \showarticletitle{Hedgecut: Maintaining randomised trees for low-latency machine unlearning}. In \bibinfo{booktitle}{\emph{Proceedings of the 2021 International Conference on Management of Data}}. \bibinfo{pages}{1545--1557}.
\newblock


\bibitem[Stoyanovich et~al\mbox{.}(2022)]%
        {stoyanovich2022responsible}
\bibfield{author}{\bibinfo{person}{Julia Stoyanovich}, \bibinfo{person}{Serge Abiteboul}, \bibinfo{person}{Bill Howe}, \bibinfo{person}{HV Jagadish}, {and} \bibinfo{person}{Sebastian Schelter}.} \bibinfo{year}{2022}\natexlab{}.
\newblock \showarticletitle{Responsible data management}.
\newblock \bibinfo{journal}{\emph{Commun. ACM}} \bibinfo{volume}{65}, \bibinfo{number}{6} (\bibinfo{year}{2022}), \bibinfo{pages}{64--74}.
\newblock


\bibitem[Tan et~al\mbox{.}(2016)]%
        {gru4rec}
\bibfield{author}{\bibinfo{person}{Yong~Kiam Tan}, \bibinfo{person}{Xinxing Xu}, {and} \bibinfo{person}{Yong Liu}.} \bibinfo{year}{2016}\natexlab{}.
\newblock \showarticletitle{Improved Recurrent Neural Networks for Session-based Recommendations}. In \bibinfo{booktitle}{\emph{Proceedings of the 1st Workshop on Deep Learning for Recommender Systems}} (Boston, MA, USA) \emph{(\bibinfo{series}{DLRS 2016})}. \bibinfo{publisher}{Association for Computing Machinery}, \bibinfo{address}{New York, NY, USA}, \bibinfo{pages}{17–22}.
\newblock
\showISBNx{9781450347952}
\urldef\tempurl%
\url{https://doi.org/10.1145/2988450.2988452}
\showDOI{\tempurl}


\bibitem[Triantafillou et~al\mbox{.}(2024)]%
        {triantafillou2024makingprogressunlearningfindings}
\bibfield{author}{\bibinfo{person}{Eleni Triantafillou}, \bibinfo{person}{Peter Kairouz}, \bibinfo{person}{Fabian Pedregosa}, \bibinfo{person}{Jamie Hayes}, \bibinfo{person}{Meghdad Kurmanji}, \bibinfo{person}{Kairan Zhao}, \bibinfo{person}{Vincent Dumoulin}, \bibinfo{person}{Julio~Jacques Junior}, \bibinfo{person}{Ioannis Mitliagkas}, \bibinfo{person}{Jun Wan}, \bibinfo{person}{Lisheng~Sun Hosoya}, \bibinfo{person}{Sergio Escalera}, \bibinfo{person}{Gintare~Karolina Dziugaite}, \bibinfo{person}{Peter Triantafillou}, {and} \bibinfo{person}{Isabelle Guyon}.} \bibinfo{year}{2024}\natexlab{}.
\newblock \bibinfo{title}{Are we making progress in unlearning? Findings from the first NeurIPS unlearning competition}.
\newblock
\newblock
\showeprint[arxiv]{2406.09073}~[cs.LG]
\urldef\tempurl%
\url{https://arxiv.org/abs/2406.09073}
\showURL{%
\tempurl}


\bibitem[Wang and Schelter(2022)]%
        {wang2022efficientlymaintainingbasketrecommendations}
\bibfield{author}{\bibinfo{person}{Benjamin~Longxiang Wang} {and} \bibinfo{person}{Sebastian Schelter}.} \bibinfo{year}{2022}\natexlab{}.
\newblock \bibinfo{title}{Efficiently Maintaining Next Basket Recommendations under Additions and Deletions of Baskets and Items}.
\newblock
\newblock
\showeprint[arxiv]{2201.13313}~[cs.IR]
\urldef\tempurl%
\url{https://arxiv.org/abs/2201.13313}
\showURL{%
\tempurl}


\bibitem[Wu et~al\mbox{.}(2023b)]%
        {Wu_2023}
\bibfield{author}{\bibinfo{person}{Jiancan Wu}, \bibinfo{person}{Yi Yang}, \bibinfo{person}{Yuchun Qian}, \bibinfo{person}{Yongduo Sui}, \bibinfo{person}{Xiang Wang}, {and} \bibinfo{person}{Xiangnan He}.} \bibinfo{year}{2023}\natexlab{b}.
\newblock \showarticletitle{GIF: A General Graph Unlearning Strategy via Influence Function}. In \bibinfo{booktitle}{\emph{Proceedings of the ACM Web Conference 2023}} \emph{(\bibinfo{series}{WWW ’23})}. \bibinfo{publisher}{ACM}, \bibinfo{pages}{651–661}.
\newblock
\urldef\tempurl%
\url{https://doi.org/10.1145/3543507.3583521}
\showDOI{\tempurl}


\bibitem[Wu et~al\mbox{.}(2023a)]%
        {kun_wu_2023}
\bibfield{author}{\bibinfo{person}{Kun Wu}, \bibinfo{person}{Jie Shen}, \bibinfo{person}{Yue Ning}, \bibinfo{person}{Ting Wang}, {and} \bibinfo{person}{Wendy~Hui Wang}.} \bibinfo{year}{2023}\natexlab{a}.
\newblock \showarticletitle{Certified Edge Unlearning for Graph Neural Networks}. In \bibinfo{booktitle}{\emph{Proceedings of the 29th ACM SIGKDD Conference on Knowledge Discovery and Data Mining}} (Long Beach, CA, USA) \emph{(\bibinfo{series}{KDD '23})}. \bibinfo{publisher}{Association for Computing Machinery}, \bibinfo{address}{New York, NY, USA}, \bibinfo{pages}{2606–2617}.
\newblock
\showISBNx{9798400701030}
\urldef\tempurl%
\url{https://doi.org/10.1145/3580305.3599271}
\showDOI{\tempurl}


\bibitem[Wu et~al\mbox{.}(2019)]%
        {srgnn}
\bibfield{author}{\bibinfo{person}{Shu Wu}, \bibinfo{person}{Yuyuan Tang}, \bibinfo{person}{Yanqiao Zhu}, \bibinfo{person}{Liang Wang}, \bibinfo{person}{Xing Xie}, {and} \bibinfo{person}{Tieniu Tan}.} \bibinfo{year}{2019}\natexlab{}.
\newblock \showarticletitle{Session-Based Recommendation with Graph Neural Networks}.
\newblock \bibinfo{journal}{\emph{Proceedings of the AAAI Conference on Artificial Intelligence}} \bibinfo{volume}{33}, \bibinfo{number}{01} (\bibinfo{date}{Jul.} \bibinfo{year}{2019}), \bibinfo{pages}{346--353}.
\newblock
\urldef\tempurl%
\url{https://doi.org/10.1609/aaai.v33i01.3301346}
\showDOI{\tempurl}


\bibitem[Wu et~al\mbox{.}(2020a)]%
        {wu2020deltagradrapidretrainingmachine}
\bibfield{author}{\bibinfo{person}{Yinjun Wu}, \bibinfo{person}{Edgar Dobriban}, {and} \bibinfo{person}{Susan~B. Davidson}.} \bibinfo{year}{2020}\natexlab{a}.
\newblock \bibinfo{title}{DeltaGrad: Rapid retraining of machine learning models}.
\newblock
\newblock
\showeprint[arxiv]{2006.14755}~[cs.LG]
\urldef\tempurl%
\url{https://arxiv.org/abs/2006.14755}
\showURL{%
\tempurl}


\bibitem[Wu et~al\mbox{.}(2020b)]%
        {Wu_2020}
\bibfield{author}{\bibinfo{person}{Yinjun Wu}, \bibinfo{person}{Val Tannen}, {and} \bibinfo{person}{Susan~B. Davidson}.} \bibinfo{year}{2020}\natexlab{b}.
\newblock \showarticletitle{PrIU: A Provenance-Based Approach for Incrementally Updating Regression Models}. In \bibinfo{booktitle}{\emph{Proceedings of the 2020 ACM SIGMOD International Conference on Management of Data}} \emph{(\bibinfo{series}{SIGMOD/PODS ’20})}. \bibinfo{publisher}{ACM}, \bibinfo{pages}{447–462}.
\newblock
\urldef\tempurl%
\url{https://doi.org/10.1145/3318464.3380571}
\showDOI{\tempurl}


\bibitem[Xia et~al\mbox{.}(2023)]%
        {simrec}
\bibfield{author}{\bibinfo{person}{Lianghao Xia}, \bibinfo{person}{Chao Huang}, \bibinfo{person}{Jiao Shi}, {and} \bibinfo{person}{Yong Xu}.} \bibinfo{year}{2023}\natexlab{}.
\newblock \showarticletitle{Graph-less Collaborative Filtering}. In \bibinfo{booktitle}{\emph{Proceedings of the ACM Web Conference 2023}} \emph{(\bibinfo{series}{WWW ’23})}. \bibinfo{publisher}{ACM}, \bibinfo{pages}{17–27}.
\newblock
\urldef\tempurl%
\url{https://doi.org/10.1145/3543507.3583196}
\showDOI{\tempurl}


\bibitem[Yu et~al\mbox{.}(2020)]%
        {dnntsp}
\bibfield{author}{\bibinfo{person}{Le Yu}, \bibinfo{person}{Leilei Sun}, \bibinfo{person}{Bowen Du}, \bibinfo{person}{Chuanren Liu}, \bibinfo{person}{Hui Xiong}, {and} \bibinfo{person}{Weifeng Lv}.} \bibinfo{year}{2020}\natexlab{}.
\newblock \showarticletitle{Predicting Temporal Sets with Deep Neural Networks}. In \bibinfo{booktitle}{\emph{Proceedings of the 26th ACM SIGKDD International Conference on Knowledge Discovery \& Data Mining}} \emph{(\bibinfo{series}{KDD ’20})}. \bibinfo{publisher}{ACM}, \bibinfo{pages}{1083–1091}.
\newblock
\urldef\tempurl%
\url{https://doi.org/10.1145/3394486.3403152}
\showDOI{\tempurl}


\bibitem[Zangerle et~al\mbox{.}(2014)]%
        {nowp}
\bibfield{author}{\bibinfo{person}{Eva Zangerle}, \bibinfo{person}{Martin Pichl}, \bibinfo{person}{Wolfgang Gassler}, {and} \bibinfo{person}{G\"{u}nther Specht}.} \bibinfo{year}{2014}\natexlab{}.
\newblock \showarticletitle{\#nowplaying Music Dataset: Extracting Listening Behavior from Twitter}. In \bibinfo{booktitle}{\emph{Proceedings of the First International Workshop on Internet-Scale Multimedia Management}} (Orlando, Florida, USA) \emph{(\bibinfo{series}{WISMM '14})}. \bibinfo{publisher}{Association for Computing Machinery}, \bibinfo{address}{New York, NY, USA}, \bibinfo{pages}{21–26}.
\newblock
\showISBNx{9781450331579}
\urldef\tempurl%
\url{https://doi.org/10.1145/2661714.2661719}
\showDOI{\tempurl}


\end{thebibliography}


\appendix

\section{Background on Machine Unlearning}
\label{sec:background}

Machine unlearning is the process of removing the influence of specific data from a machine learning model as if that data had never been seen during training. For the formal definition of unlearning consider the following: Let $D$ be the original training dataset, $\mathcal{A}$ be a training algorithm, $U$ be an unlearning algorithm, $D_f \subseteq D$ be the forget set to unlearn, and $D_r = D \setminus D_f$ be the retain set, e.g. the data retained after unlearning. $U$ is an exact unlearning algorithm iff for all $T \subseteq \mathcal{H}$:
$$
    \mathbb{P}(\mathcal{A}(D_r) \in T) = \mathbb{P}(U(\mathcal{A}(D), D_r, D_f) \in T)
$$
where $\mathcal{H}$ is the hypothesis space of models. The intuition behind the formula is that the probability of receiving a model $\mathcal{A}(D_r)$ after retraining from scratch on the retain set $D_r$ should be the same as the probability of receiving this model from the unlearning procedure $U$ starting at the model $\mathcal{A}(D)$ unlearning $D_f$. This implies the same distribution of weights and model outputs.
For theoretical guarantees for approximate unlearning there is the notion of $(\varepsilon, \delta)$-unlearning~\cite{triantafillou2024makingprogressunlearningfindings}, where $U$ is a $(\varepsilon, \delta)$-unlearning algorithm iff for all $T \subseteq \mathcal{H}$:
$$
\mathbb{P}(\mathcal{A}(D_r) \in T) \leq \exp(\varepsilon) \, \mathbb{P}(U(\mathcal{A}(D), D_r, D_f) \in T) + \delta
$$
and
$$
\mathbb{P}(U(\mathcal{A}(D), D_r, D_f) \in T) \leq \exp(\varepsilon) \, \mathbb{P}(\mathcal{A}(D_r) \in T) + \delta
$$
with $\varepsilon \in [0, \infty), \delta \in [0, 1)$. Here, $\varepsilon$ is a multiplicative error margin, and $\delta$ is an additive error margin, where lower values of these parameters denote stronger unlearning bounds. Note that $(0, 0)$-unlearning is equivalent to exact unlearning.
\section{Additional unlearning algorithms} \label{app:unlearning-algorithms}
\begin{itemize}[noitemsep,leftmargin=*]
    \item \texttt{GIF}, which works similarly to SCIF, but the parameters considered for the parameter update are chosen differently: The node embedding of node $u$ in the graph is contained in the parameter update if there exists a node $v$ such that $v$ is in the forget set and the distance from $u$ to $v$ is smaller than a hyperparameter $d$~\cite{Wu_2023}.
    \item \texttt{CEU}, a framework to unlearn edges from a GNN. To get theoretical guarantees for linear models they first add a linear noise term $b^T \theta$ with $b \sim \mathcal{N}(0, \sigma^2 I)$ to the training loss and fine-tune aiming to hide the real gradient residual. After that, they unlearn using an influence function on all parameters~\cite{kun_wu_2023}.
    \item \texttt{IDEA}, which analyses the original loss over the whole training set and the pruned loss which excludes nodes and edges we want to unlearn. It then conducts a second-order update step based on the difference between the original and pruned objective loss~\cite{dong_2024}.
    \item \texttt{Caboose}, an exact unlearning algorithm for neighborhood-based recommendation on sparse data, which uses efficient indexing and incremental view maintenance to update nearest neighbor sets with sub-second latency~\cite{schelter2023forget,schelter2024snapcase}. 
\end{itemize}

\end{document}